\documentclass{aa}  
\usepackage{pgfplots}
\usepackage{graphicx}
\usepackage{txfonts}
\usepackage{mathrsfs}
\usepackage[colorlinks,allcolors=blue]{hyperref}
\usepackage{breakurl}
\usepackage{subfigure}

\def \src {\mbox{A0538$-$66}}
\def \ero {\emph{eROSITA}}
\def \xmm {\emph{XMM-Newton}}
\begin{document}

   \title{\emph{eROSITA} detection of flares from the Be/X-ray binary A0538$-$66}
\titlerunning{\emph{eROSITA} detection of flares from A0538$-$66}

   \author{L. Ducci
          \inst{1,2}
          \and
          S.Mereghetti
          \inst{3}
          \and
          A. Santangelo
          \inst{1}
          \and
          L. Ji
          \inst{1}
          \and
          S. Carpano
          \inst{4}
          \and
          S. Covino
          \inst{5}
          \and
          V. Doroshenko
          \inst{1}
          \and
          F. Haberl
          \inst{4}
          \and
          C. Maitra
          \inst{4}
          \and
          I.~Kreykenbohm
          \inst{6}
          \and
          A. Udalski
          \inst{7}
   }

   \institute{Institut f\"ur Astronomie und Astrophysik, Kepler Center for Astro and Particle Physics, Universit\"at T\"ubingen, 
              Sand 1, 72076 T\"ubingen, Germany\\
              \email{ducci@astro.uni-tuebingen.de}
              \and
              ISDC Data Center for Astrophysics, Universit\'e de Gen\`eve, 16 chemin d'\'Ecogia, 1290 Versoix, Switzerland
              \and
              INAF -- Istituto di Astrofisica Spaziale e Fisica Cosmica, Via A. Corti 12, 20133 Milano, Italy
              \and
              Max-Planck-Institut für extraterrestrische Physik, Gießenbachstraße 1, 85748 Garching, Germany
              \and
              INAF -- Osservatorio Astronomico di Brera, via Bianchi 46, 23807 Merate (LC), Italy
              \and
              Universit\"at Erlangen/N\"urnberg, Dr.-Remeis-Sternwarte, Sternwartstraße 7, D-96049, Bamberg, Germany
              \and
              Astronomical Observatory, University of Warsaw Al. Ujazdowskie 4 00-478 Warszawa, Poland
   }

   \date{Received ...; accepted ...}

   \abstract
       {In 2018, \xmm\ observed the awakening in X-rays of the Be/X-ray binary (Be/XRB) \src.
        It showed bright and fast flares close to periastron with properties that had never been observed in other Be/XRBs before.
        We report the results from the observations of \src\
        collected during the first all-sky survey of \ero, an X-ray telescope (0.2$-$10\,keV)
        on board the Spektrum-Roentgen-Gamma (SRG) satellite.
        \ero\ caught two flares within one orbital cycle at orbital phases
        $\phi = 0.29$ and $\phi = 0.93$ (where $\phi=0$ corresponds to periastron),
        with peak luminosities of $\sim 2-4 \times 10^{36}$\,erg\,s$^{-1}$ (0.2$-$10\,keV) and
        durations of $42 \leq \Delta t_{\rm fl} \leq 5.7\times 10^4$\,s.
        The flare observed at $\phi \approx 0.29$ shows that the neutron star can accrete considerably
        far from periastron, although it is expected to be outside of the circumstellar disk,
        thus providing important new information about the plasma environment surrounding the binary system.
        We also report the results from the photometric monitoring of \src\ carried out with
        the REM, OGLE, and MACHO telescopes from January 1993 until March 2020. 
        We found that the two sharp peaks that characterize the orbital modulation in the optical
        occur asymmetrically in the orbit, relative to the position of the donor star.
       }


 \maketitle

\section{Introduction} 
\label{sect intro}

\src\ is a Be/X-ray binary (Be/XRB) in the Large Magellanic Cloud (LMC).
It hosts a B1\,IIIe star \citep{Rajoelimanana17} and a neutron star (NS)
whose 69\,ms pulsation was detected only once, during a super-Eddington X-ray outburst
($L_{\rm x} \approx 8\times 10^{38}$\,erg\,s$^{-1}$ in 1.1$-$21\,keV; \citealt{Skinner82}).
The orbital period is $P_{\rm orb}\approx16.64$\,d and the eccentricity is $e\approx 0.72$
\citep[][and references therein]{Rajoelimanana17}.
The lack of detection of pulsations in all the other outbursts,
whose luminosities ranged from $\sim 10^{34}$\,erg\,s$^{-1}$ to $\sim 10^{38}$\,erg\,s$^{-1}$,
led \citet{Campana95} and \citet{Corbet97} to propose that the accretion onto
the NS surface (which produces X-ray pulsations) is only possible when the mass
capture rate is very high. When the luminosity is lower, the plasma cannot
overcome the centrifugal barrier, and accretion is thus inhibited.
\citet{Skinner82} and \citet{Campana95} estimated an upper limit
for the magnetic dipole moment of $\mu\lesssim 10^{29}$\,G\,cm$^3$ for A0538-66.
In 2018, after a long period in which the source showed a faint X-ray emission
or  was barely detected\footnote{for example, in an \xmm\ observation
carried out in 2002, \src\ showed an X-ray luminosity of
$L_{\rm x} \approx 5-8 \times 10^{33}$\,erg\,s$^{-1}$ in 0.3-10\,keV \citep{Kretschmar04}.},
\xmm\ observed its awakening in two observations carried out near periastron:
a remarkable flaring activity consisting of
fast flares ($\sim$2$-$50\,s) that reached peak luminosities up to
$\sim 4\times 10^{38}$\,erg\,s$^{-1}$ (0.2$-$10\,keV).
Between the flares, the X-ray luminosity
was $\sim 2\times 10^{35}$\,erg\,s$^{-1}$.
In a third observation at periastron, \xmm\ detected \src\ at a constant
luminosity level of $\sim 2\times 10^{34}$\,erg\,s$^{-1}$ \citep{Ducci19a}.
During these flares, the X-ray spectrum showed two distinct components.
The softer component ($\lesssim 2$\,keV) was well described by an absorbed power law
with photon index $\Gamma_{\rm soft}\approx 2-4$ and column density $N_{\rm H}\approx 10^{21}$\,cm$^{-2}$.
The harder component ($\gtrsim 2$\,keV) was described by a power law with
photon index $\Gamma_{\rm hard}\approx 0-0.5$.
The softer component showed larger flux variability than the harder component
and a hardening correlated with the flux.
No statistically significant pulsations were detected.
A flaring activity with these properties has not been observed in other Be/XRBs.
Most of them show a weak X-ray emission ($L_{\rm x}\lesssim 10^{34}-10^{35}$\,erg\,s$^{-1}$)
interrupted by periodic or sporadic outbursts (typically $L_{\rm x}\lesssim 10^{38}$\,erg\,s$^{-1}$)
that last from several days to weeks.
The X-ray outbursts are caused by the
accretion of plasma from the circumstellar disk of the Be star onto the NS \citep[see][for a review]{Reig11}.
It is argued that the activation mechanism of the X-ray outbursts
and their main properties (duration and peak luminosity) depend on the interactions between
the compact object and a warped, eccentric, and truncated
circumstellar disk \citep{Laplace17, Martin11, Martin14, Okazaki02}.

\src\ also shows a striking variability in the optical, characterized by
periodic outbursts associated with the X-ray outbursts.
They are the brightest optical outbursts observed in Be/XRBs.
The brightest outbursts ($\Delta m_{\rm v} \approx 2.2$; \citealt{Charles83})
are likely powered by the X-ray photons produced by the accreting NS,
which are reprocessed in an envelope surrounding the binary system.
For the faintest flares, the production mechanism is unknown:
it could be the same as for the bright flares, although
the heating of the surface of the donor star irradiated
by the NS or other mechanisms cannot be excluded \citep[][and references therein]{Ducci19b}.
Photometric data from 1915 to 1998, based on UV Schmidt and Harvard photographic plates
and Massive Compact Halo Objects (MACHO) data, revealed a long-term
modulation of $P_{\rm sup}=420.8 \pm 0.8$\,d, with a reddening at low fluxes
and the lack of the $\sim  16.64$\,d flares and of the H$\alpha$ emission line
during the high-luminosity state of the $\sim420$\,d cycle.
The long-term variability was explained with the formation and depletion of a circumstellar
disk observed edge-on \citep{Alcock01, McGowan03, Ducci16, Rajoelimanana17}.
\citet{Schmidtke14} analyzed the Optical Gravitational Lensing Experiment (OGLE\,IV)
observations and showed that since 2010, the long-term variability began to occur on irregular timescales.
This new behavior was later observed with the Rapid Eye Mount (REM) telescope \citep{Ducci16}.
The MACHO and OGLE light curves folded
at the orbital period show two sharp peaks around periastron.
The lack of an eclipse implies that the orbital plane is not observed edge-on,
and it must thus have a substantial inclination
with respect to the circumstellar disk \citep{Rajoelimanana17}.

We report the results of the analysis of the data of \src\
from the first \ero\ all-sky survey.
\src\ was detected in X-rays and surprisingly showed an X-ray flare
far from periastron. No optical data are available during the X-ray brightening detected by \ero.
Despite this, a new analysis of the MACHO and OGLE data, enriched with the data from a daily
monitoring specifically tailored for this source performed with
REM (a large part of which we present here for the first time),
shows other important properties of \src.

\section{Data analysis}
\label{sect data analysis}

\subsection{Optical photometric data}
\label{sect optical photometric data}

We used the available MACHO, OGLE, and REM photometric data of \src.
We retrieved the MACHO data\footnote{\url{http://macho.anu.edu.au} \src\ identified as 61.9045.32 in the MACHO catalogue.} that cover the time interval 49001.6$-$51542.6 MJD (14 January 1993$-$30 December 1999),
and we extracted the photometric light curve in the $V$ and $R$ bands using the calibration
procedure described in \citet{Alcock99} and \citet{Alcock01}.
We retrieved the OGLE data\footnote{\url{http://ogle.astrouw.edu.pl/ogle4/xrom/xrom.html}},
corrected to the standard $I$ -band system \citep{Udalski08}.
The OGLE data cover the time interval 55260.1$-$58921.1 MJD (5 March 2010$-$13 March 2020).
We also used the REM photometric data in the $g$, $r$, $i$ bands.
REM is a fully automatic, fast-reacting telescope
  with a 60\,cm primary mirror at the ESO La Silla Observatory \citep{Covino04,Zerbi04}.
  It hosts the ROSS2 optical camera \citep[$g$, $r$, $i$, $z$; ][]{Tosti04}
  and the REMIR near-infrared camera \citep[$J$, $H$, $K$; ][]{Conconi04}.
  REM allows observing a target in five bands simultaneously (one in the near-infrared band,
  and four in the optical band). 
REM data provide a long-term monitoring of this source
that consists of a denser sampling (almost daily monitoring)
than the MACHO and OGLE data sets,
and is thus ideal for studying the fast variability shown by \src.
These data cover the time intervals 56903.2$-$57223.4\,MJD (3 September 2014$-$20 July 2015)
and 57477.0$-$58929.3\,MJD (30 March 2016$-$21 March 2020).
For data reduction methods, we refer to section 2 in \citet{Ducci16},
where the results from the first part of the data were presented.

\subsection{eROSITA}

The extended ROentgen Survey with an Imaging Telescope Array (\ero)
is the soft X-ray instrument on board the Russian/German Spektrum-Roentgen-Gamma (SRG) mission,
launched in July 2019 and placed in orbit around the $L_2$ Lagrangian point \citep{Merloni12}.
\ero\ consists of seven identical mirror modules combined with as many
\emph{pn}-CCDs operating in the 0.2$-$10\,keV energy range.
It has an angular resolution of $\lesssim 15^{\prime\prime}$ and a field of view of $1.03^{\circ}$.
In December 2019, \ero\ started a four-year-long all-sky survey,
composed of eight passages over the entire sky. \ero\ scans are regular, with an average
uniform exposure of $\sim 150-200$\,s over most of the sky.
One of the two ecliptic poles is close to
the position of the LMC.
This region is thus observed more frequently, and a total exposure of up to a few hours is reached \citep{Predehl20}.
Additionally, the LMC was observed a few days prior to the start of the first survey
 as part of a test observations program.
Therefore \ero\ observed the region around
\src\ several times, from 8 December 2019 at 16:01:12 (UTC)
until 7 June 2020 at 16:58:48 (UTC),
corresponding to the time interval 58825.6675$-$59007.7075\,MJD,
for a net exposure time of $\sim 4666$\,s.

We reduced the data with the \ero\ Standard Analysis Software System (eSASS),
version 201009, pipeline configuration c946.
We did not use data from the telescope modules without on-chip
filters for blocking optical light (TM5 and TM7) because their energy calibration is unreliable.
First, we ran the source detection in the energy range 0.2$-$10\,keV,
adopting a minimum detection likelihood of $L=6$.
(the detection likelihood is defined as $L=-\ln(p)$, where $p$ is the
probability that a Poissonian function in the background is detected
as a spurious source).
\src\ was clearly detected with a likelihood of $L\approx 250$.

\begin{figure}[h!]
\begin{center}
  \includegraphics[width=8.6cm]{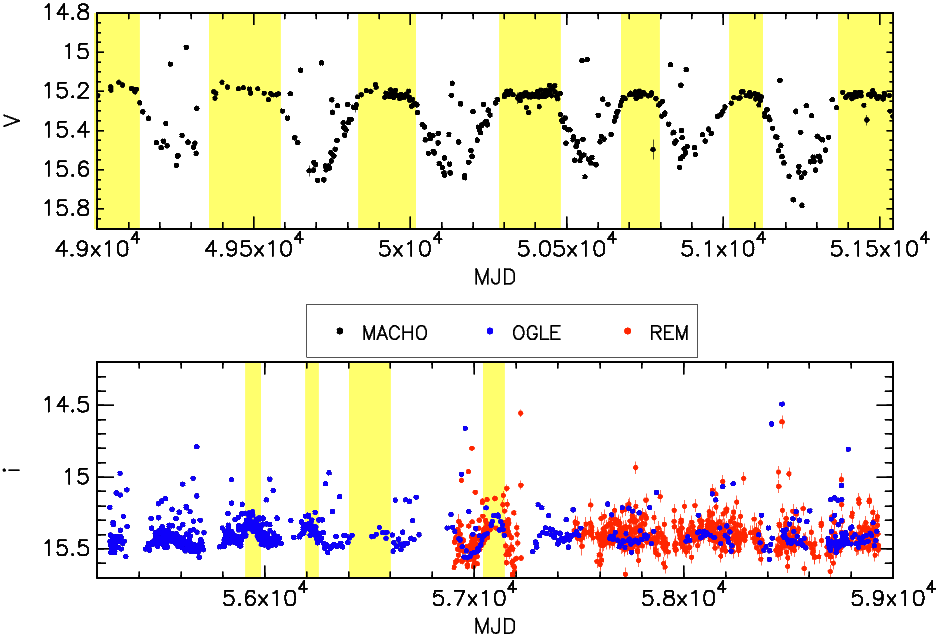}
  \includegraphics[width=8.1cm]{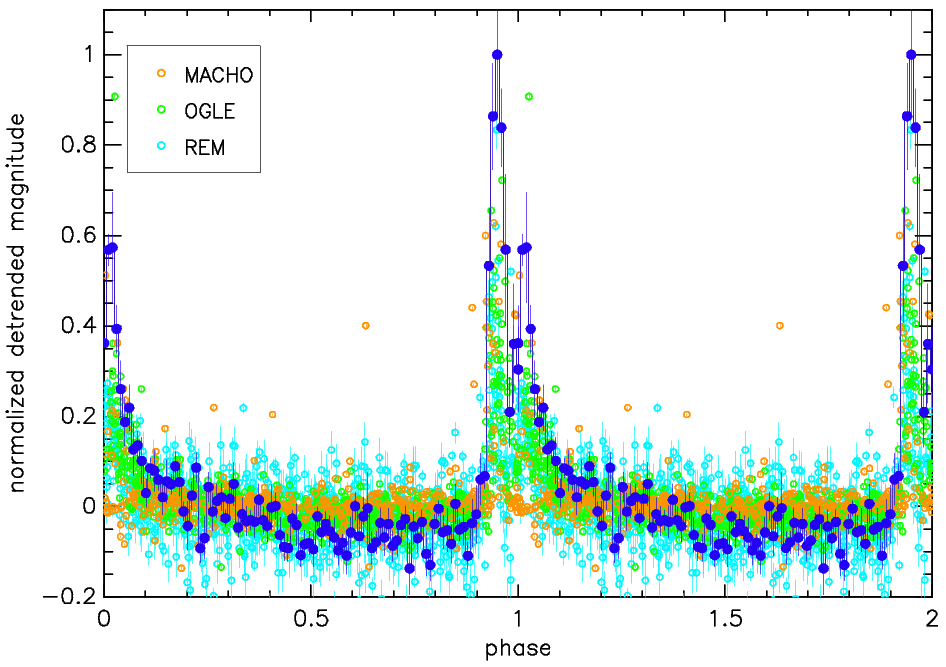}\\
  \includegraphics[width=8.5cm]{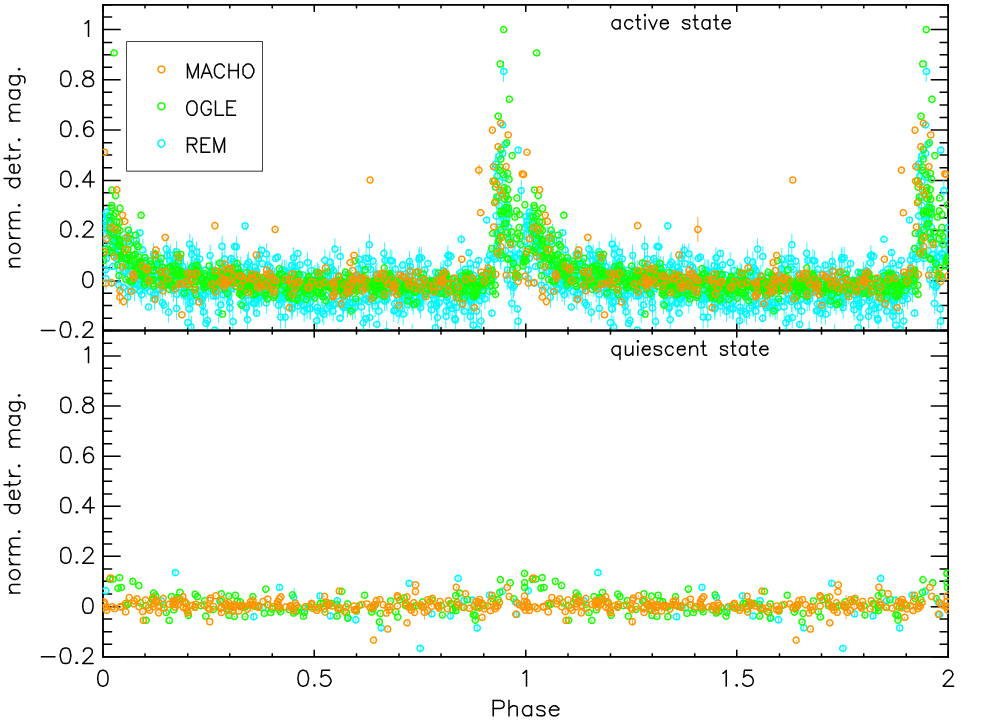}\\
\end{center}
\caption{\emph{Top two panels:} MACHO, OGLE, and REM light curves of \src. Yellow regions show the quiescent states, when the flaring activity is absent \citep{Alcock01,McGowan03}.
\emph{Central panel:} Same data as in the upper panel, folded at the orbital period ($P_{\rm orb} =  16.64002$\,d).
Phase zero corresponds to the periastron passage (according to the orbital ephemeris obtained by \citealt{Rajoelimanana17}). Filled blue circles show the binned light curve. Colored empty circles are the individual photometric points.
  \emph{Bottom panel:} MACHO, OGLE, and REM light curves of \src\ folded at the orbital period, divided into two groups:
  1) Time intervals corresponding to the active state (i.e., with flaring activity and circumstellar disk) and
  2) time intervals corresponding to the quiescent state (see yellow regions in the top panel). 
  When error bars are not visible, they are smaller than the plotted symbols.}
\label{figREM}
\end{figure}

\begin{figure}
\begin{center}
  \includegraphics[width=8cm]{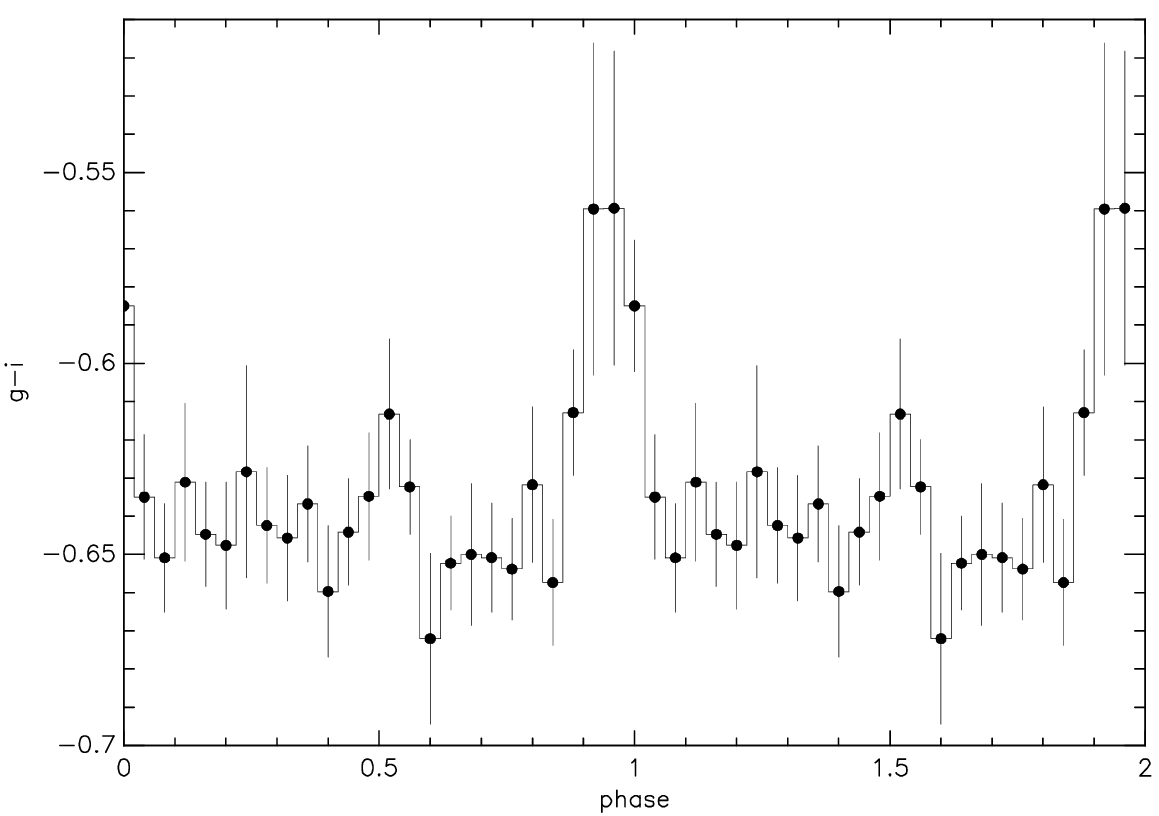}
\end{center}
\caption{$g-i$ color from REM data taken from 57477.0 to 58929.3\,MJD
  (i.e., the second part of the data, where \src\ did not show any long-term variability),
  folded at the orbital period of \src. Phase zero corresponds to the periastron passage.}
\label{figcolor}
\end{figure}

\section{Results}
\label{sect results}

\subsection{Optical data}

We removed the long-term modulation in the optical data and obtained
the detrended light curve as follows.
In each data set (MACHO, OGLE, and REM), we temporarily removed the fast periodic flares ($\sim16.6$\,d).
Then, we divided each data set into smaller segments of length $\sim 100-150$\,d.
Each segment was fit with a third-order polynomial.
The best-fit functions were then subtracted from the original light curves.
The orbital modulation of \src\ is not sinusoidal \citep[e.g.,][]{Rajoelimanana17}.
Therefore we searched for periodicity using the phase dispersion minimization (PDM)
periodogram \citep{Stellingwerf78}, which is better suited for this type of signal.
To obtain a better measurement of the periodicity and to search for possible orbital decay,
we again divided the detrended light curve into five subsets: 1) MACHO, 2) OGLE part 1 (55260.100$-$57090.613\,MJD), 3) OGLE part 2 (57090.613$-$58921.000\,MJD), and 4) REM part 1 and REM part 2 (see Sect. \ref{sect optical photometric data}).
Then, we performed a phase-coherent timing analysis,
using a linear and a quadratic function \citep[e.g.,][]{DallOsso03, Falanga15}.
We find that the linear function is sufficient to obtain a statistically acceptable solution
for the ephemeris of the optical outburst ($\chi^2=1.15$, 3\,degrees of freedom):
$T_0=55673.71\pm0.05$\,MJD (position of the first peak), and $P_{\rm orb}=16.64002\pm0.00026$\,d.
The second peak falls $\Delta \phi \approx 0.07$ after the first. 
The top two panels of Fig. \ref{figREM}
  display the MACHO, OGLE, and REM light curves.
  The yellow regions show time intervals without flaring activity, called ``quiescent state''
by \citet{Alcock01} and \citet{McGowan03}, as opposed to the ``active states'', when \src\ shows X-ray and optical flares.
The light curves of these two panels show the vanishing of the superorbital period,
which  might be explained with a circumstellar disk that becomes more stable over the years
in which the depletion events are more sporadic.
The third panel shows 
the binned light curve (filled blue circles) and
the individual points (empty colored circles) folded at the orbital period.
Phase zero corresponds to the periastron passage \citep[see the orbital ephemeris in ][]{Rajoelimanana17}.
The lower panel shows the optical light curves folded at the orbital period
divided between active and quiescent states.
The difference between the phases of the peaks reported in this work and in \citet{Rajoelimanana17}
arises because we used a larger data set, which enabled us to obtain a more accurate ephemeris.

Bright flares
are clustered in the narrow orbital phase range $-0.075 \lesssim \phi \lesssim 0.075$,
with the exception of one point at phase $\phi \approx 0.64$.
It corresponds to an event detected by MACHO on 50160.67\,MJD (18 March 1996).
No X-ray observations correspond to this event. We also note that the MACHO observations
before and after this event were taken on 50154.6\,MJD and 50173.5\,MJD, when \src\ was faint in the optical.
Because this interval is quite long, it is possible that the bright event detected on 50160.67\,MJD
was part of a longer (up to $\sim 20$\,d) optical and X-ray outburst, similar to the outbursts observed in 1970s-1980s.
The profile shown in Fig. \ref{figREM} is reminiscent of the fast rise-exponential decay (FRED) profiles seen in many other Be/XRBs \citep{Bird12} with the difference that \src\ shows a double peak (or, in an alternative interpretation, a dip at the position of the expected peak).

\begin{figure}
\begin{center}
  \includegraphics[width=8cm]{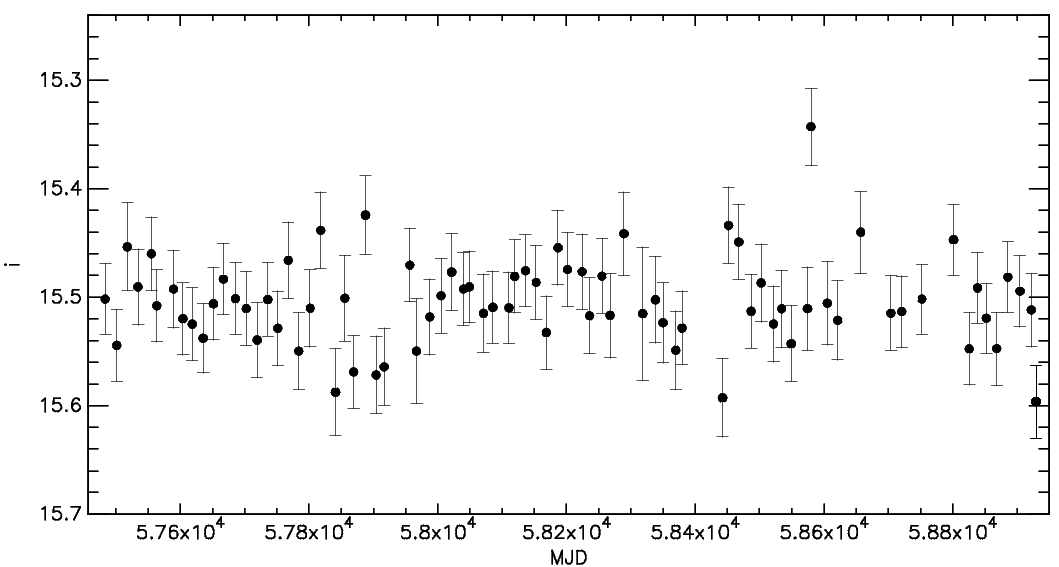}
\end{center}
\caption{REM light curve of \src\ from March 2016 until March 2020. The bin size corresponds to one orbital period.}
\label{fig:long-termREM}
\end{figure}

Figure \ref{figcolor} shows the $g-i$ color folded on the orbital period of \src,
built using the second part of the REM data. This subset was ideal
to study the $g-i$ variability because in addition to the $\sim 16.6$\,d orbital variability,
it does not show the long-term modulation previously seen, especially in MACHO data
(see the light curve of the recent REM monitoring of \src\ in Fig. \ref{fig:long-termREM}).
Figure \ref{figcolor} shows that \src\ is redder at high luminosities.

\begin{figure*}
\begin{center}
  \includegraphics[height=8cm]{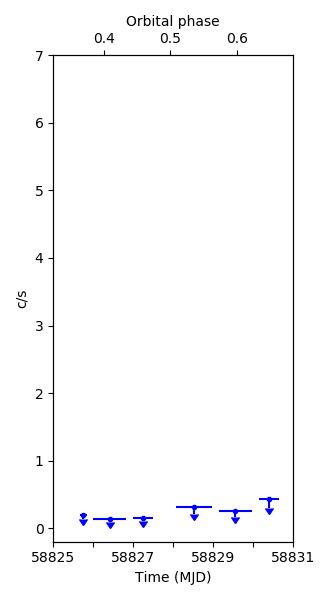}
  \includegraphics[height=8cm]{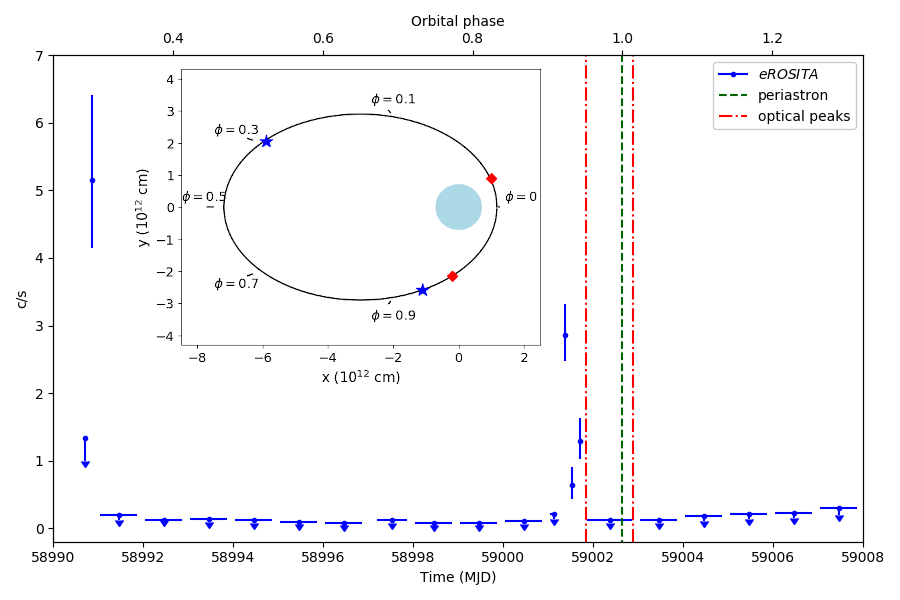}
\end{center}
\caption{\ero\ light curve (0.2-10 keV) of \src\ of the first sky survey. Each point during the flares represents one scan. Outside of the flares, the 90\% c.l. upper limit was obtained by binning the data within one day.
  The top horizontal axis shows the orbital phase, and phase zero corresponds to the periastron passage.
  The inset shows the position of the \ero\ flares (blue stars) and of the optical peaks (red squares) in the orbit.
}
\label{figlcr}
\end{figure*}

\subsection{\ero}

Figure \ref{figlcr} shows the 0.2$-$10\,keV light curve of \src.
It is built using events extracted from a circular region
with radius $50^{\prime\prime}$ centered on the source position (RA: 05:35:41.00 Dec: $-$66 51 53.5).
Background counts were accumulated from a circular region with J2000 coordinates
RA: 05:36:15.39 Dec: $-$66:52:29.6 radius: $106^{\prime\prime}$,
inside which we excluded a circular region with coordinates
RA: 05:32:07.65 Dec: $-$66:52:30.4 and radius $12^{\prime\prime}$,
where a faint source was detected.

\src\ is detected by \ero\ in four observation scans
at $\sim 58990.87$\,MJD and $\sim 59001.54$\,MJD
(orbital phases $\sim 0.29$ and $\sim 0.93$).
The upper and lower error bars of the flares were calculated using the
Gehrels approximation (equations 9 and 14 in \citealt{Gehrels86}).
The 90\% confidence level (c.l.) upper limits
were obtained by binning observations within one day.

The flare at $\phi \sim 0.29$ (called flare 1, see Table \ref{Table1})
has a duration between 42.5\,s and 28759.5\,s (data gaps prevent a more accurate estimate).
Flare 2 ($\phi \sim 0.92$) was observed in three \ero\ observation scans.
It has a duration between 28857.0\,s and 57542.2\,s.
The observation scans containing the flares do not have enough statistics to allow
a meaningful spectral analysis (the source counts for flares\,1 and \,2  are 90, and there are 15 background counts).
To obtain a rough estimate of the absorbed luminosity,
we made use of the \ero\ response file. Then, for the spectral shape, we assumed the best-fit model obtained in \citet{Ducci19a} for the intermediate
state of the flaring activity seen by \xmm. We obtain
$L_{\rm f,1}= 4.2 \pm 1.1 \times 10^{36}$\,erg\,s$^{-1}$ and
$L_{\rm f,2}= 2.3 \pm 0.4 \times 10^{36}$\,erg\,s$^{-1}$ (0.2$-$10\,keV).
Outside of the flares, \src\ is neither detected in single observations
(the typical $3\sigma$ upper limit for the single-observation scans is $\approx 2\times 10^{35}$\,erg\,s$^{-1}$)
nor in the bins obtained by combining different observation scans, as shown in Fig. \ref{figlcr}.
A source detection performed in all the observations taken outside the flares
combined together does not find any source at the position of \src.
Assuming the best-fit model
of the low state seen by \xmm\ \citep{Ducci19a} for the spectral shape, we obtain a $3\sigma$
upper limit for the absorbed luminosity of
$L_{\rm u.l.}=3.4 \times 10^{34}$\,erg\,s$^{-1}$ (0.2$-$10\,keV).

\begin{table*}
  \begin{center}
    \caption{Summary tables of the flares and low-luminosity level of \src\ seen by \ero.}
    \label{Table1}
      \begin{tabular}{lcccccc}
        \hline
        \hline
        \noalign{\smallskip}
        & $\phi_{\rm orb}$ &          Time                    & $\Delta t_{\rm exp}$\,$^a$ & $\Delta t_{\rm min}$\,$^b$ &  $\Delta t_{\rm max}$\,$^c$ & $L_{\rm x}$\,$^d$ \\
        \noalign{\smallskip}
        &                &              (MJD)               &         (s)             &          (s)             &            (s)            &   (erg\,s$^{-1}$)       \\
        \noalign{\smallskip}
        \hline
        \noalign{\smallskip}        
Flare 1 &     0.29       &              58990.873           &        21       &         42.5             &          28759.5          & $4.2 \pm 1.1\times 10^{36}$ \\
        \noalign{\smallskip}
Flare 2 &     0.93       &              59001.540           &        121      &       28857.0            &          57542.2          & $2.3 \pm 0.4\times 10^{36}$ \\
        \noalign{\smallskip}
Outside of the flares &  &       58825.667$-$59007.707      &      4524       &                      &                            & $\leq 3.4\times 10^{34}$ ($3\sigma$ u.l.) \\
        \noalign{\smallskip}
        \hline
      \end{tabular}
  \end{center}
  Notes. $^a$: Net exposure.
  $^b$: Minimum duration of the flare, based on the observation scans where the source is detected.
  $^c$: Upper limit on the duration of the flares, based on observations before and after those of the flare detections.
  $^d$: Absorbed luminosity in the energy range $0.2-10$\,keV, assuming a distance of 50\,kpc.
\end{table*}

\section{Discussion}
\label{sect discussion}

\begin{figure}
\begin{center}
  \includegraphics[width=9cm]{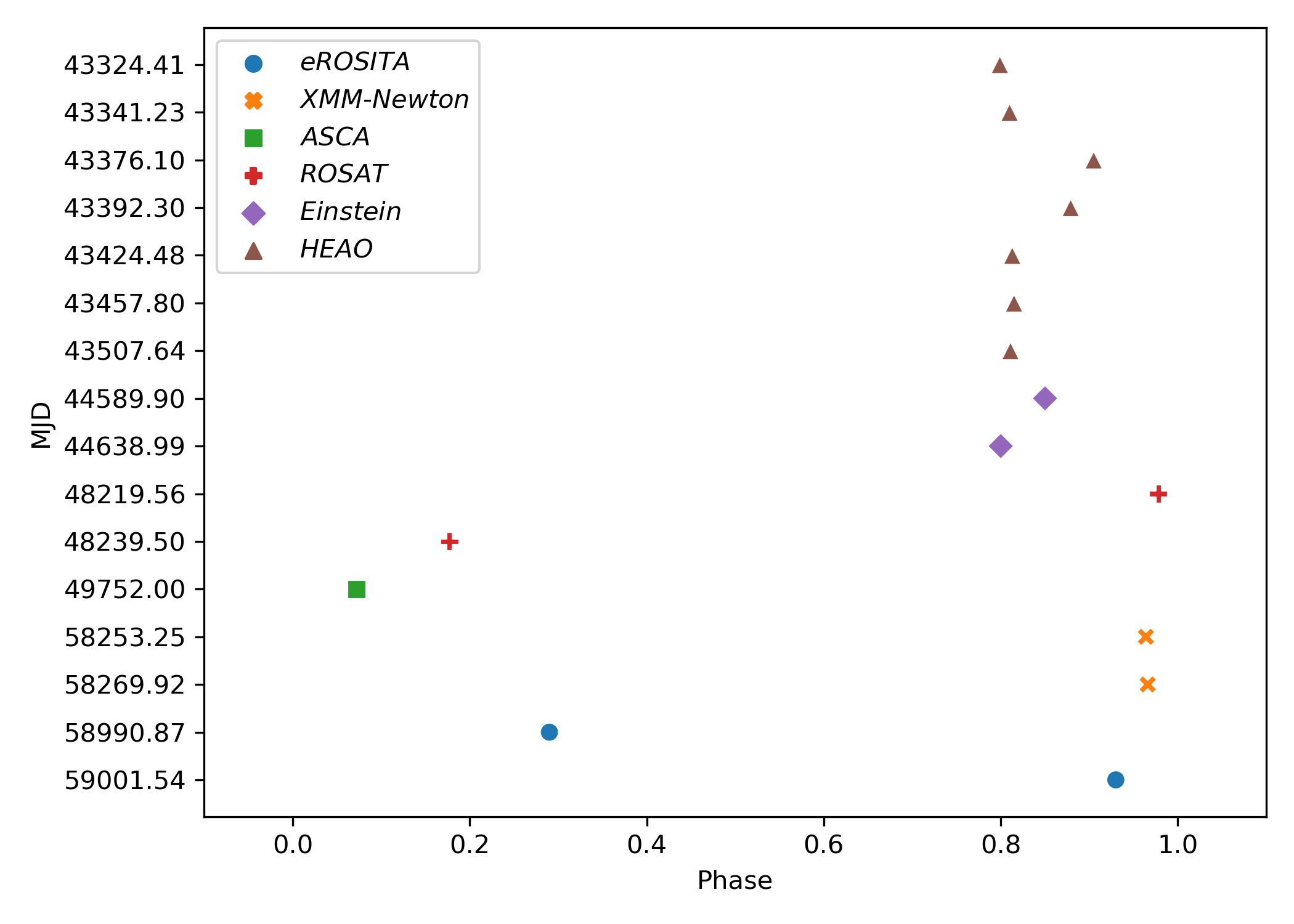}
\end{center}
\caption{Phases of the observed X-ray flares of \src\ shown in Table \ref{Table2}.}
\label{fig:xpeaks}
\end{figure}

As described in Sect. \ref{sect intro}, X-ray and optical data of \src\ suggest that the NS orbit and the circumstellar
disk are highly misaligned and have two intersections close to the periastron.
When the NS is outside of the circumstellar disk, it accretes the fast and
tenuous polar wind of the Be star, which leads to low mass accretion rates (the mass-loss rate and terminal velocity of the polar wind
of Be stars are $\dot{M}\approx 10^{-10}-10^{-8}$\,M$_\odot$\,yr$^{-1}$ and $v_\infty\approx 600-1800$\,km\,s$^{-1}$; e.g. \citealt{Waters88}).
When the NS crosses the circumstellar disk, where the outflow from the Be star is denser
and slower (in the circumstellar disk $\dot{M} \sim 100-1000$ higher and $v_\infty \sim 10-100$
lower than in the polar wind region; e.g. \citealt{Waters88}), the higher accretion rate produces
X-ray flares, and consequently, optical flares from reprocessing of X-ray photons \citep{Ducci19a}.
There are three points emerging from the observations which are difficult
to reconcile with this scenario.
First: the flare 1 detected by \ero.
Second:  Table \ref{Table2} and Fig. \ref{fig:xpeaks} show a list of all the X-ray outbursts of \src\
  observed since its discovery with the orbital phases of the observed peaks.
  The list shows that the peaks of some outbursts preceeded the expected epochs for the optical flares.
Third: Fig. \ref{figlcr} shows that the orbital phases in which the peaks of the two optical flares occur are
  not compatible with a scenario in which the NS crosses a circumstellar disk twice that entirely
  lies on the equatorial plane of the Be star. If this had occured, the orbital phases of the flares would be aligned with the center of the Be star.

A highly disturbed environment surrounding the binary system
could explain these observational features.
Hydrodynamic simulations by \citet{Martin14,Martin11} showed that in
binary systems in which the orbital plane and circumstellar
disk are misaligned, the disk can become warped and eccentric.
The variability of the H$\alpha$ emission profile during the orbital
  cycle of \src\ observed by \citet{Rajoelimanana17} supports
  the hypothesis that the circumstellar disk can be warped.
  In addition, flare 1 observed by \ero\ and the other flares seen in the past (Table \ref{Table2})
  show that the NS might occasionally encounter a structure of gas resulting
from tidal interactions at any orbital phase
and produce enhanced X-ray emission also far from periastron.
The reddening at high luminosities (Fig. \ref{figcolor}) suggests that
optical flares are associated with a brightening of a cold region around the binary system
(e.g., produced by a sudden increase in the size of a cold emitting region, likely the circumstellar disk, caused by the tidal displacement of material during periastron passages of the NS).
It is unclear whether both optical flares are equally affected by the reddening. More precise observations are critical to clarify this point.
We point out that the reddening at high luminosities
during the orbital cycle shown in Fig. \ref{figcolor} should not be
confused with the reddening at low fluxes during the super-orbital cycle
mentioned in Sect. \ref{sect intro}, which is caused by the
formation of a circumstellar disk observed edge-on.

The recent high X-ray emission observed by \ero\ and \xmm\
is different from that observed in other high-mass X-ray binaries with Be and OB supergiant donor stars (see the discussion in \citealt{Ducci19a}). In particular, the X-ray flares of \src\ are remarkably shorter than the X-ray outbursts displayed by the other Be/XRBs,
which last for a large fraction of the orbit ($\approx 0.2-0.3$\,$P_{\rm orb}$)
or for several orbital cycles \citep[e.g.,][]{Reig11}.
For most of these Be/XRBs, there is compelling evidence for disk-mediated accretion.
Accretion disks were also suggested in the context of the hydrodynamic simulations
of \citet{Martin14} to explain the properties of type II outbursts
produced by the mass capture from an eccentric and misaligned circumstellar disk.
For \src, long periods of accretion from a spherically symmetric flow seem a more plausible scenario
\citep{Ducci19a}.
Nonetheless, if an accretion disk of the type of those hypothesized by \citet{Syunyaev77} forms around the NS of \src\ and persists for several orbits, a stochastic leakage of matter onto the NS surface could also explain the flare observed by \ero\ at $\phi \approx 0.29$.

To our knowledge, the Be/XRB that most resembles \src\ for its X-ray and optical variability is AX\,J0049.4$-$7323.
It is located in the Small Magellanic Cloud and hosts a $\sim 750$\,s
pulsar with an orbital period of $\sim 393$\,d \citep{Yokogawa00,Cowley03,Laycock05}.
AX\,J0049.4$-$7323 shows periodic bright ($L_{\rm x}\gtrsim 10^{37}$\,erg\,s$^{-1}$)
and short ($0.05-0.1$\,$P_{\rm orb}$) X-ray outbursts that are synchronized with optical outbursts that last as long as the X-ray ourbursts,
have optical amplitudes of $\Delta m_I\approx 0.5,$ and are characterized by
a fast rise-exponential decay profile (figure 2 in \citealt{Ducci19c}).
They occur at the same orbital phase, likely at periastron.
Except for \src, these are the brightest optical outbursts
associated with the orbital period seen in Be/XRBs.
Moreover, AX\,J0049.4$-$7323 reddens during the flares,
possibly because of the brightening of the circumstellar disk
during the outbursts \citep{Ducci19c,Cowley03}.
Another similarity with \src\ is that AX\,J0049.4$-$7323 sometimes shows substantial X-ray variability (by a factor of $\sim 270$)
with peak luminosities of $\approx 2\times 10^{36}$\,erg\,s$^{-1}$
far from periastron \citep[][and references therein]{Ducci18}.
\citet{Coe04} pointed out that two other very little studied Be/XRBs,
RX\,J0058.2$-$7231 and RX\,J0520.5$-$6932,
show signs of highly disturbed circumstellar disks.
These four sources could thus be the members
of a noteworthy subclass of Be/XRBs.

\begin{table}
  \begin{center}
    \caption{Summary table of all the observed X-ray flares of \src\ since its discovery.}
    \label{Table2}
    \resizebox{\columnwidth}{!}{
      \begin{tabular}{lccccc}
        \hline
        \hline
        \noalign{\smallskip}
        Phase    &   MJD    & $\Delta t$   & Satellite & Reference \\
        \hline
        \noalign{\smallskip}
        0.799    & 43324.41 & 6\,hr        & \emph{HEAO} & \citet{Skinner80} \\
        0.810    & 43341.23 & 2\,hr        & \emph{HEAO} & \citet{Skinner80} \\
        0.905    & 43376.10 & 14\,d        & \emph{HEAO} & \citet{Skinner80} \\
        0.879    & 43392.30 & 3\,d         & \emph{HEAO} & \citet{Skinner80} \\
        0.813    & 43424.48 & 0.6\,d       & \emph{HEAO} & \citet{Skinner80} \\
        0.815    & 43457.80 & 0.1-0.6\,d   & \emph{HEAO} & \citet{Skinner80} \\
        0.811    & 43507.64 & $\approx$0.1\,d & \emph{HEAO} & \citet{Skinner80} \\

        0.850$^b$& 44589.90 & unknown      & \emph{Einstein} & \citet{Ponman84} \\
        0.800$^b$& 44638.99 & unknown      & \emph{Einstein} & \citet{Ponman84} \\

        0.978    & 48219.56 & 12\,hr       & \emph{ROSAT} & \citet{Mavromatakis93} \\
        0.177$^a$& 48239.50 & 10\,d        & \emph{ROSAT} & \citet{Mavromatakis93} \\

        0.072$^c$ & 49752.00 & $>0.6$\,d    & \emph{ASCA} & \citet{Corbet97} \\
        0.964     & 58253.25 & $\approx 1$\,hr & \emph{XMM-Newton} & \citet{Ducci19a} \\
        0.966     & 58269.92 & $\approx 1$\,hr & \emph{XMM-Newton} & \citet{Ducci19a} \\
        0.290     & 58990.87 & $\geq 42.5$\,s & \ero\ & this work \\
        0.930     & 59001.54 & $\geq 8$\,hr& \ero\ & this work \\
        \noalign{\smallskip}
        \hline
      \end{tabular}
    }
  \end{center}
  Notes. $^a$ Onset of the outburst at $\phi_{\rm orb}\approx 0$\\
  $^b$ $\Delta t$ unknown. Outburst could be much longer than a few days\\
  $^c$ MACHO data show a simultaneous optical outburst with a peak between 49749.578 and 49751.651 ($\phi_{\rm orb}=0.927-0.051$)
\end{table}

\section{Conclusion}
\label{sect conclusion}

\ero\ confirmed the anomalous flaring activity of \src\ previously seen by \xmm\ in 2018, and it detected an unexpected flare
far from periastron.
We propose that this flare and the other observational features emerging from optical data
and previous X-ray observations could be qualitatively explained if the binary system
were surrounded by an inhomogeneous environment, likely a warped
circumstellar disk, possibly with noncoplanar structures of gas detached from the disk through tidal interactions with the NS.
This hypothesis is mostly based on hydrodynamic simulations by \citet{Martin14}
for a binary system with an orbital period of $\sim 24$ days and a moderate eccentricity ($e=0.34$),
and it is supported by the spectroscopic observations of \src\ reported
in \citet{Rajoelimanana17}.
Although the results of the simulations of \citet{Martin14} are generally applicable to other Be/XRBs,
the extreme properties of \src\ (e.g., it has the shortest orbital period and the highest
eccentricity of all Be/XRBs, as well as a very high misalignment between the circumstellar disk and the orbital plane)
require further hydrodynamic simulations that are specifically tailored for this source.
We also point out that a growing number of Be/XRBs with similar properties might benefit from this study.

\begin{acknowledgements}
  We  thank  the  anonymous  referee  for  constructive  comments that helped to improve the paper.
  This work is partially supported by the \textsl{Bundesministerium f\"{u}r Wirtschaft und Energie} through the \textsl{Deutsches Zentrum f\"{u}r Luft- und Raumfahrt e.V. (DLR)} under the grants FKZ 50 QR 2102 and DLR 50 QR 2103.
  SM acknowledges   support through grants ASI-INAF n.2017-14-H.0, MIUR 2017LJ39LM “UNIAM” and INAF PRIN-SKA/CTA.
  The OGLE project has received funding from the National Science Centre, Poland, grant MAESTRO 2014/14/A/ST9/00121 to AU.
  This work is based on data from eROSITA, the soft X-ray instrument aboard SRG, a joint Russian-German science mission supported by the Russian Space Agency (Roskosmos), in the interests of the Russian Academy of Sciences represented by its Space Research Institute (IKI), and the Deutsches Zentrum für Luft- und Raumfahrt (DLR).
  This paper utilizes public domain data obtained by the MACHO Project, jointly funded by the US Department of Energy through the University of California, Lawrence Livermore National Laboratory under contract No. W-7405-Eng-48, by the National Science Foundation through the Center for Particle Astrophysics of the University of California under cooperative agreement AST-8809616, and by the Mount Stromlo and Siding Spring Observatory, part of the Australian National University.
\end{acknowledgements}

\bibliographystyle{aa} 
\bibliography{lducci_eRosita_A0538}

\end{document}